# Graph Neural Network for Predicting the Effective Properties of Polycrystalline Materials: A Comprehensive Analysis

Minyi Dai[1], Mehmet F. Demirel[2], Xuanhan Liu[2], Yingyu Liang[2], Jia-Mian Hu[1*]

[1]Department of Materials Science and Engineering, University of Wisconsin-Madison, Madison, WI, 53706, USA

[2]Department of Computer Sciences, University of Wisconsin-Madison, Madison, WI, 53706, USA

## Abstract

We develop a polycrystal graph neural network (PGNN) model for predicting the effective properties of polycrystalline materials, using the $Li_7La_3Zr_2O_{12}$ ceramic as an example. A large-scale dataset with >5000 different three-dimensional polycrystalline microstructures of finite-width grain boundary is generated by Voronoi tessellation and processing of the electron backscatter diffraction images. The effective ion conductivities and elastic stiffness coefficients of these microstructures are calculated by high-throughput physics-based simulations. The optimized PGNN model achieves a low error of <1.4% in predicting all three diagonal components of the effective Li-ion conductivity matrix, outperforming a linear regression model and two baseline convolutional neural network models. Sequential forward selection method is used to quantify the relative importance of selecting individual grain (boundary) features to improving the property prediction accuracy, through which both the critical and unwanted node (edge) feature can be determined. The extrapolation performance of the trained PGNN model is also investigated. The transfer learning performance is evaluated by using the PGNN model pretrained for predicting conductivities to predict the elastic properties of the same set of microstructures.

*Corresponding author: jhu238@wisc.edu

# 1. Introduction

Establishing the microstructure-property relationship lies at the core of materials science and engineering. A microstructure is typically represented by a 3D image comprised of discretized voxels, where each voxel stores the physical features of interest such as the crystallographic orientation, concentration, magnetization, and polarization. The effective property of a material is the average of the responses of these local physical features to an applied field [1]. Such local response is critically determined by the physical fields (e.g., stress, magnetic, and electric field) arising due to the interactions of the local physical features stored in different voxels. It is therefore necessary to consider these physical interactions for realizing an accurate prediction of the effective material properties [2].

Machine learning (ML) has recently emerged as a powerful tool to predict the microstructure-property relationship [3–5]. One key reason is that the prediction by a ML model can be orders of magnitude faster than physics-based modeling especially for a large-scale 3D microstructure image that contain billions of voxels [6–8]. Different types of ML models have been developed to predict the microstructure-property relationship [9–28]. Among them, the convolutional neural network (CNN) based method [13–25,28] is particularly attractive for two main reasons. First, it directly operates on the image voxels and hence is applicable to any types of microstructures. Second, the influence of neighboring voxels on one voxel is incorporated during the convolution process, which enables parameterizing the local physical interactions. However, for polycrystalline materials, the influence of neighboring grains and grain boundaries on a given grain, which is critical to the effective properties [29–31], cannot be incorporated. This is because both grains and grain boundaries cannot be processed as individual units in image-voxel-based representation.

Graph-based ML models, such as the graph neural network (GNN), have recently been applied to predict the effective properties of polycrystalline materials [32–39]. In a GNN model, a graph — a set of interacting nodes that are connected by edges — is used to represent a polycrystalline microstructure which contains a set of interacting grains that are connected by grain boundaries. Compared to CNN, GNN has two unique capabilities in predicting the microstructure-property relationship. First, graph enables a more compact representation of a polycrystalline microstructure because one grain typically comprises multiple voxels. As a result, the training time of a GNN model is usually much shorter than that of CNN models [34,39]. Second, the message passing function of a GNN model enables incorporating both the local and nonlocal physical interactions among the grains or grain boundaries. Therefore, it is rational to expect that a GNN model can outperform a CNN model in predicting the effective properties of polycrystalline materials.

Despite this expectation, the property prediction performance of GNN has not yet been systematically evaluated against baseline CNN models in existing reports [32–39], except one single study where the GNN only shows marginal improvement against a VGGNet CNN model [39]. Moreover, all these GNN models employ training datasets where physics-based simulations of effective properties were performed by omitting the contribution from the grain boundaries. This makes it not possible to quantify the importance of the physical features of grain boundary in GNN-based property prediction via a feature selection [36,40] or feature attribution [34,41] study. Furthermore, in predicting the properties of molecules and atomic crystal structures, it was reported that a GNN model pretrained on a larger dataset can be transferred to improve the performance of predicting other properties of the same molecular/crystal structure with smaller training data [42]. However, such transfer learning capability of GNN has not yet been demonstrated in the case of microstructure-property prediction.

In this article, we perform high-throughput physics-based simulations to calculate the effective ion conductivity and the effective elastic stiffness tensors of 5000 different 3D polycrystalline microstructures, using $Li_7La_3Zr_2O_{12}$ ceramic as a representative material. In our physics-based simulations, grain boundary is treated as a monolithic phase that has its own thickness, crystallographic orientation, ion conductivity, and elastic stiffness tensors. Built upon the crystal graph convolutional neural network [43] developed for predicting the properties of atomic crystal structure, we develop a *Polycrystal Graph Neural Network (PGNN)* model for predicting the effective properties of 3D polycrystalline microstructures. The PGNN model is then trained, validated, and tested using the dataset we created.

It is found that the PGNN model gives a testing mean absolute relative error of < 1.4% in predicting all three diagonal components of the effective ion conductivity tensor, and this error is appreciably lower than those from a linear regression model, a 3D CNN model that has recently been utilized to predict the effective elastic moduli of two-phase composites [28], and the ResNet [44] which is a widely used general-purpose CNN model. We demonstrate a good extrapolation performance of the trained PGNN model using a separate testing dataset where the polycrystalline microstructures of finite-width grain boundary are obtained by processing of the electron backscatter diffraction (EBSD) data. Moreover, the sequential forward selection method is employed to quantify the relative importance of the different physical features of the grains and grain boundaries to the prediction accuracy, which enables identifying both the critical and unwanted node/edge features. An excellent transfer learning performance of the PGNN model is also demonstrated. Compared to a PGNN model trained from scratch, a PGNN model pretrained using dataset of microstructures and effective ion conductivities shows a significantly shorter training time and lower error in predicting the effective elastic stiffness coefficients of the same set of microstructures.

## 2. Methods

*2.1. Building a microstructure graph*

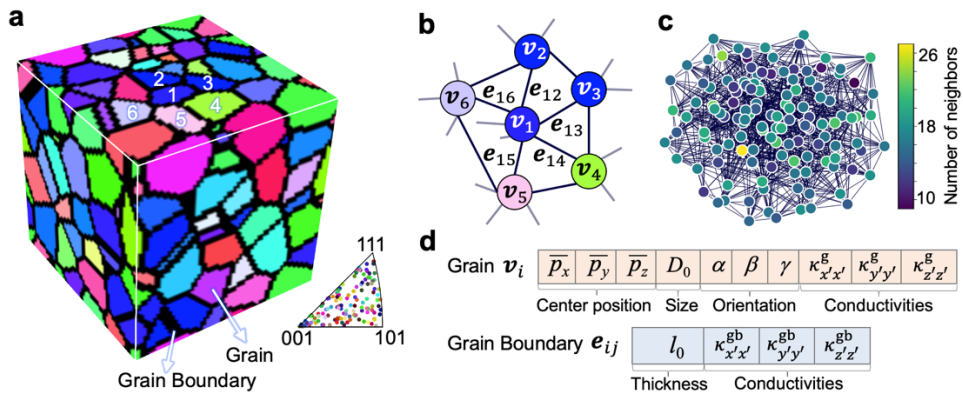

**Figure 1.** (**a**) A typical 3D polycrystalline microstructure image in the dataset. There are 133 grains in this specific microstructure. Grains are colored based on their Euler angles, and the colors are indicated by the inverse pole figure. Grain boundaries are colored black. (**b**) Schematic of building an undirected microstructure graph from the microstructure image. (**c**) Microstructure graph built from the microstructure image in (a). (**d**) Summary of grain (node) features and grain boundary (edge) features.

Figure 1a shows a representative 3D polycrystalline microstructure generated via an in-house Voronoi Tessellation code, where each grain is colored according to its crystallographic orientation defined by the three Euler angles ($\alpha$, $\beta$, $\gamma$) and the grain boundary is colored black. The microstructure is not textured, as shown by the inverse pole figure. To build a graph for this microstructure, each grain and each grain boundary are labeled and represented by a node and an edge, respectively. For each node and edge, there is an associated vector (denoted as $\boldsymbol{v}_i$ and $\boldsymbol{e}_{ij}$) which stores the physical features of the corresponding grain and grain boundary, respectively. The subscript of the vectors ($i,j$=1,2,3…,$N$) represent the label of a specific grain, where $N$ is the total number of grains in one microstructure. For example, $e_{12}$ refers to the grain boundary that connects the two grains which are labeled as '1' and '2', as shown by Fig. 1b. Thus, the whole microstructure can be converted into an undirected graph with nodes and edges. Figure 1c shows the full microstructure graph, where each node is colored based on the number of edges connected to it (*i.e.*, the number of neighboring grains).

The target properties are the effective ion conductivities $\kappa_{xx}^{\text{eff}}$, $\kappa_{yy}^{\text{eff}}$, $\kappa_{zz}^{\text{eff}}$ or the effective elastic stiffness coefficients $c_{11}^{\text{eff}}$, $c_{12}^{\text{eff}}$, $c_{44}^{\text{eff}}$. In this regard, the physical features in the vectors $\boldsymbol{v}_i$ and $\boldsymbol{e}_{ij}$ should include the ion conductivity or elastic moduli of individual grains or grain boundaries. Otherwise, the application of ML to predict the effective conductivities or moduli could fall into the pitfall of finding correlations that do not necessarily exist. With this in mind, we select ten components for each node feature vector $\boldsymbol{v}_i$, including three coordinates for the grain center position, one for the grain size, three for the grain orientation, and three for the local ion conductivity. Specifically, since each voxel has its unique coordinates ($p_x$, $p_y$, $p_z$) as well as a label $n_i$ indicating that the voxel belongs to the $i^{\text{th}}$ grain, the coordinates of the center of the $i^{\text{th}}$ grain can be calculated by averaging the coordinates of all the voxels labeled as $n_i$ under the periodic boundary condition. The grain size ($D_0$) is defined as the number of voxels occupied by a grain. The grain orientation is represented by the three Euler angles ($\alpha$, $\beta$, $\gamma$). The local ion conductivity is represented by the three diagonal components of the conductivity matrix $\kappa_{x'x'}^{\text{g}}$, $\kappa_{y'y'}^{\text{g}}$, $\kappa_{z'z'}^{\text{g}}$, where the superscript 'g' refers to the grain and the prime symbols in the subscripts indicate that these properties are in the crystallographic coordinate system. For the edge feature vector, we select four components, including the grain boundary thickness ($l_0$) and conductivity $\kappa_{x'x'}^{\text{gb}}$, $\kappa_{y'y'}^{\text{gb}}$, $\kappa_{z'z'}^{\text{gb}}$. The necessity of selecting these physical features as node/edge features will be evaluated based on feature selection study (Section 4.4).

*2.2. Graph input, network structure and update function of the PGNN*

Based on the node and edge feature vectors $\boldsymbol{v}_i$ and $\boldsymbol{e}_{ij}$ for each microstructure graph, three matrices are constructed as the input of the PGNN, including an $N\times10$ node feature matrix **F** that stores the $v_i$ of each node, an $N\times N\times4$ edge feature matrix **E** that stores the $\boldsymbol{e}_{ij}$ of each edge, an $N\times N$ adjacency matrix **A** that stores the adjacency relation (where $A_{ij}$=1 if grain $i$ and grain $j$ are neighbors and $A_{ij}$=0 otherwise). Figure 2 illustrates these three input matrices and the four main layers of the PGNN model. The four layers are described below.

*(1) The embedding layer.* The embedding layer takes the node feature matrix $F$ as the input and converts the original node features into node embeddings $F^0$ through a linear transformation,

$$\mathbf{F}_e^0 = \mathbf{F}\mathbf{W}_e + \mathbf{b}_e \tag{1}$$

where $\mathbf{W}_e$ is a weight matrix and $\mathbf{b}_e$ is the bias term. Both are trainable. This embedding layer allows for transferring the remaining layers to learn other datasets having different data dimensions.

*(2) The graph convolutional layer.* Multiple graph convolutional layers can be used. The first graph convolutional layer takes three matrices as its input: the node embedding matrix $\mathbf{F}_g^0 = \mathbf{F}_e^0$, the adjacency matrix $\mathbf{A}$, and the edge feature matrix $\mathbf{E}$. The graph convolutional layer updates the features of all the nodes through the same update function, while the matrix $\mathbf{A}$ and $\mathbf{E}$ remain unchanged in this process. Specifically, after passing the $n$th graph convolution layer ($n$=1,2,3,…), the node embedding matrix $\mathbf{F}_g^{n-1}$ is updated to $\mathbf{F}_g^n$ through,

$$\mathbf{F}_g^n = ReLU\left[\left(\mathbf{A}\mathbf{F}_g^{n-1} \oplus \sum_{i=1}^{N} \mathbf{E}[:,i,:] \oplus \mathbf{F}_g^{n-1}\right)\mathbf{W}_g^n + \mathbf{b}_g^n\right], \quad (2)$$

where $\sum_{i=1}^{N} \mathbf{E}[:,i,:]$ is the sum of the features of all edges connected to the $i^{th}$ node, which changes the matrix dimension from $N \times N \times 4$ to $N \times 4$; the symbol $\oplus$ means matrix concatenation; $\mathbf{W}_g^n$ is the trainable weight matrix of the $n^{th}$ graph convolutional layer; $\mathbf{b}_g^n$ is the bias term of the $n^{th}$ graph convolutional layer; $ReLU(x) = max(0,x)$ is an activation function. Equation (2) is the matrix form of the update function of node features [43], i.e.,

$$v_i^n = ReLu\left[\left(\sum_j v_i^{n-1} \oplus e_{ij}\right)\mathbf{W}_c^n + v_i^{n-1}\mathbf{W}_s^n + \mathbf{b}^n\right], \quad (3)$$

Where $v_i^n$ and $v_i^{n-1}$ are the feature vector of $i^{th}$ node in the $n^{th}$ and $(n-1)^{th}$ layer, respectively; $e_{ij}$ is the feature vector of edge $(i, j)$; $\mathbf{W}_c^n$ and $\mathbf{W}_s^n$ are the trainable weight matrices of the $n^{th}$ graph convolutional layer; $\mathbf{b}^n$ is the bias term of the $n^{th}$ graph convolutional layer. Equation (3) indicates that the updated features of a specific node are determined by the features of all the edges connected to it as well as the features of all its neighboring nodes. By utilizing multiple graph convolutional layers, the interactions among all grains and grain boundaries can be considered.

*(3) The node-level fully connected layer.* Multiple node-level fully layers can be used. The first node-level fully connected layer takes the node embedding matrix generated by the last graph convolutional layer as the input, that is, $\mathbf{F}_f^0 = \mathbf{F}_g^n$. After passing the $m$th layer, the node embedding matrix $\mathbf{F}_f^{m-1}$ is updated to $\mathbf{F}_f^m$ through,

$$\mathbf{F}_f^m = softplus(softplus(\mathbf{F}_f^{m-1})\mathbf{W}_f^m + \mathbf{b}_f^m), \quad (4)$$

where the $\mathbf{W}_f^m$ is the trainable weight matrix of the $m^{th}$ node-level fully connected layer; $\mathbf{b}_f^m$ is the bias term of the $m^{th}$ node-level fully connected layer; $softplus(x) = log(1 + exp(x))$ is utilized as the nonlinear activation function, but the $ReLU$ can also be used. Such node-wise dense layers have previously been used in GNN models developed for predicting the properties of atomic crystal structures and molecules [43,45], and the goal of introducing such layers is to provide more trainable parameters while maintain scalability with respect to the number of nodes.

*(4) Fully connected layer.* The node embedding matrix generated by the node-level fully connected layer is flattened to a one-dimensional vector $\mathbf{f}^0$ and used as the input of the fully connected layer. After passing the $l^{th}$ layer, the vector $\mathbf{f}^{l-1}$ is updated to $\mathbf{f}^l$ through,

$$\mathbf{f}^l = ReLU(\mathbf{f}^{l-1}\mathbf{W}_f^l + \mathbf{b}^l), \tag{5}$$

where $\mathbf{W}_f^l$ is the trainable weight matrix of the $l^{th}$ layer and $\mathbf{b}^l$ is the bias term of the $l^{th}$ layer. In the last fully connected layer which is linked to the target properties, the $ReLU$ activation function is removed to enable the prediction of negative targets.

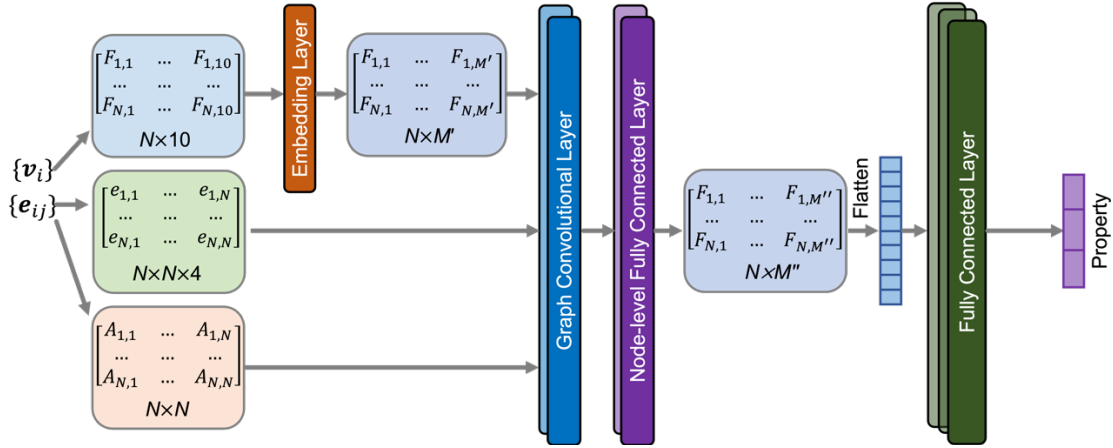

**Figure 2.** The network structure of the PGNN model.

## 3. Datasets Generation

To evaluate the property prediction performance of the PGNN model against the baseline machine learning models, a dataset with 5000 data points of 3D polycrystalline microstructures and their corresponding effective ion conductivities are generated. Specifically, each data point is denoted as $\{\mathbf{X}, \mathbf{Y}\}$, where $\mathbf{X}$ is the microstructure input and has a dimension of $N_x \times N_y \times N_z \times 3$, where $N_x=N_y=N_z=64$ are the number of voxels along the three Cartesian axes and the additional three channels store the three Euler angles; $\mathbf{Y}$ refers to the effective ion conductivities ($\kappa_{xx}^{\text{eff}}$, $\kappa_{yy}^{\text{eff}}$, $\kappa_{zz}^{\text{eff}}$). In this dataset, 4000, 500, and 500 data points are utilized for model training, validation, and testing, respectively. To evaluate the transfer learning performance of the PGNN model, a smaller dataset of 604 data points is generated, where each data point is denoted as $\{\mathbf{X}, \mathbf{Y}'\}$. Here the microstructure input of 604 different $\mathbf{X}$ is randomly picked from the 4000 microstructures in the training dataset, and the $\mathbf{Y}'$ refers to the corresponding effective elastic stiffness coefficients ($c_{11}^{\text{eff}}$, $c_{12}^{\text{eff}}$, $c_{44}^{\text{eff}}$), which are utilized as the new target properties. We note that the effective ion conductivity and the effective elastic stiffness tensors are two key properties of the solid electrolytes for lithium batteries [46,47].

### 3.1. Generation of 3D polycrystalline microstructure of finite-width grain boundary

The 3D polycrystalline microstructures, each having a size of 64×64×64 voxels, are generated through an in-house Voronoi Tessellation code. We begin by randomly selecting $N$ voxels as the seeds of $N$ grains, and label them from 1 to $N$. For each of the remaining voxels, its distances to all the $N$ seeds are calculated. Each voxel is then given the same label as its nearest seed. As a result, the 3D cube is divided into $N$ different regions with unique labels, representing $N$ different grains. If the label of a voxel is different from those of its neighboring voxels, the voxel will be marked as the grain boundary and re-labeled as 0. Figure 3a shows the statistical distribution of the size of all the grains (top panel) and the number of grains $N$ per microstructure image (bottom panel) in the 5000 polycrystalline microstructures.

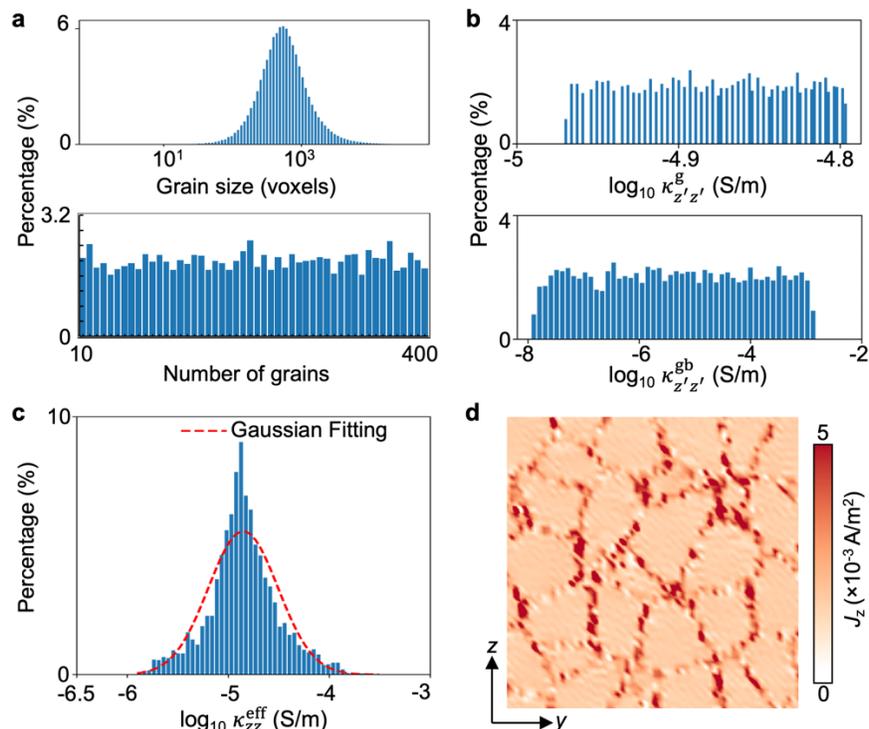

**Figure 3**. (**a**) Statistical distributions of the size of all the grains and the number of grains per microstructure in all 5000 3D polycrystalline microstructures generated by Voronoi tessellation. Statistical distributions of (**b**) the input grain conductivity $\log_{10}\kappa^{g}_{z'z'}$ and grain conductivity $\log_{10}\kappa^{gb}_{z'z'}$ (crystallographic coordinate system) and (**c**) the calculated effective Li-ion conductivities $\log_{10}\kappa^{eff}_{zz}$ (sample coordinate system) for all the 5000 microstructures. (**d**) Spatial distribution of the calculated electrical current density component $J_z$ in one 2D slice of one specific 3D microstructure.

After the generation of the grains and grain boundaries, voxels inside the same grain are given the same set of Euler angles ($\alpha$, $\beta$, $\gamma$), where the values of $\alpha$, $\beta$, $\gamma$ are randomly selected from the range of {0°, 360°}, {0°,180°}, {0°, 360°}, respectively. In different grains of a 3D microstructure, the Euler angles are different. The crystallographic orientations of grain boundaries can be characterized by five macroscopic degrees of freedoms (three parameters on crystal misorientation and two parameters on the normal axis orientation) [29]. In this work, we assign randomly selected Euler angles to all voxels that are labeled as the grain boundary. This simplified treatment can approximate the atomistic disordered of the grain boundaries in the LLZO ceramic [31]. As shown in Figs. 3b-c, although the distributions of the input local grain and grain boundary conductivities ($\kappa^{g}_{z'z'}$ and $\kappa^{gb}_{z'z'}$) are largely uniform, the distribution of the target effective conductivity ($\kappa^{eff}_{zz}$) is Gaussian. This suggests a strong nonlinear correlation between the input and output properties.

*3.2. Calculation of the effective ion conductivities and elastic stiffness coefficients*

The calculations of the effective ion conductivities **Y** and the effective elastic stiffness matrix **Y'** are performed based on high-throughput physics-based simulations via the commercial software μ-Pro® (mupro.co), which take the local Euler angles ($\alpha$, $\beta$, $\gamma$) and the local ion conductivity (or local elastic stiffness) matrix of the polycrystalline LLZO ceramic as the input.

The effective electrical conductivity $\kappa_{kl}^{\text{eff}}$ is given by $\kappa_{kl}^{\text{eff}} = \frac{\langle J_k(\mathbf{r})\rangle}{E_l}$ ($k,l=x,y,z$) where $E_l$ is the electrical field applied along the $l$ axis, and $\langle J_k(\mathbf{r})\rangle$ is the volumetric average of the $k$ component of the local electrical current density $J_k(\mathbf{r})$. The 3D spatial distribution of the $J_k(\mathbf{r})$ under the applied $E_l$ is obtained by solving the steady-state continuity equation,

$$\nabla \cdot \mathbf{J}(\mathbf{r}) = \nabla \cdot (\boldsymbol{\kappa}(\mathbf{r})\mathbf{E}) = 0 \tag{6}$$

via a numerically efficient Fourier spectral iterative perturbation (FSIP) method [48,49]. Here $\boldsymbol{\kappa}(\mathbf{r})$ is the local ion conductivity in the sample coordinate system. Assuming all the non-diagonal components are zero, $\boldsymbol{\kappa}(\mathbf{r})$ is given by,

$$\boldsymbol{\kappa}(\mathbf{r}) = \mathbf{R}\boldsymbol{\kappa}'(\mathbf{r})\mathbf{R}^T = \mathbf{R}\begin{bmatrix} \kappa_{x'x'}(\mathbf{r}) & 0 & 0 \\ 0 & \kappa_{y'y'}(\mathbf{r}) & 0 \\ 0 & 0 & \kappa_{z'z'}(\mathbf{r}) \end{bmatrix}\mathbf{R}^T, \tag{7}$$

where the rotation matrix $\mathbf{R}$ is defined based on the local Euler angles $(\alpha, \beta, \gamma)$,

$$\mathbf{R} = \begin{bmatrix} \cos\alpha\cos\gamma - \cos\beta\sin\alpha\sin\gamma & -\cos\alpha\sin\gamma - \cos\beta\cos\gamma\sin\alpha & \sin\alpha\sin\beta \\ \cos\gamma\sin\alpha + \cos\alpha\cos\beta\sin\gamma & \cos\alpha\cos\beta\cos\gamma - \sin\alpha\sin\gamma & -\cos\alpha\sin\beta \\ \sin\beta\sin\gamma & \cos\gamma\sin\beta & \cos\beta \end{bmatrix}. \tag{8}$$

$\kappa_{x'x'}$, $\kappa_{y'y'}$ and $\kappa_{z'z'}$ are the local Li-ion conductivities in the crystallographic coordinate system. In voxels that are labeled as the grain, $(\kappa_{x'x'}, \kappa_{y'y'}, \kappa_{z'z'}) = (\kappa_{x'x'}^g, \kappa_{y'y'}^g, \kappa_{z'z'}^g)$. Here we assume that the grains in the LLZO ceramic have a tetragonal crystal symmetry due to residual stress, and we denote the tetragonal axis as the $z'$ axis. If all grains have a cubic crystal symmetry, $\kappa_{x'x'}(\mathbf{r}) = \kappa_{y'y'}(\mathbf{r}) = \kappa_{z'z'}(\mathbf{r})$. In that case, the local conductivity matrix $\boldsymbol{\kappa}$ would be independent of the local grain orientations since $\boldsymbol{\kappa} = \mathbf{R}\boldsymbol{\kappa}'\mathbf{R}^T = \boldsymbol{\kappa}'\mathbf{I} = \boldsymbol{\kappa}'$, where $\mathbf{I}$ is the identity matrix. For this reason, we set $\kappa_{x'x'}^g = \kappa_{y'y'}^g = 1.335 \times 10^{-5}$ S/m, which are the reported values for LLZO ceramics [31], but further set $\kappa_{z'z'}^g = r_1 \kappa_{x'x'}^g$, where $r_1$ is randomly chosen from a uniform distribution in the range of 0.8-1.2. In voxels that are labeled as the grain boundary, $(\kappa_{x'x'}, \kappa_{y'y'}, \kappa_{z'z'}) = (\kappa_{x'x'}^{gb}, \kappa_{y'y'}^{gb}, \kappa_{z'z'}^{gb})$, where $\kappa_{x'x'}^{gb} = 10^{r_2}\kappa_{x'x'}^g$, $\kappa_{y'y'}^{gb} = 10^{r_3}\kappa_{y'y'}^g$ and $\kappa_{z'z'}^{gb} = 10^{r_4}\kappa_{z'z'}^g$. Here, $r_2$, $r_3$ and $r_4$ are randomly selected from a uniform distribution in the range of -3 and 2 because the reported conductivity of grain boundaries in LLZO can be higher or lower than the grain conductivity [31]. Importantly, among the 5000 polycrystalline microstructures, each one has a unique set of $r_1$, $r_2$, $r_3$ and $r_4$ values. Therefore, the distributions of local Li-ion conductivity in the 5000 microstructures are all different from each other. In each microstructure, the same set of $(\kappa_{x'x'}^g, \kappa_{y'y'}^g, \kappa_{z'z'}^g)$ is utilized for all the grains, and the same set of $(\kappa_{x'x'}^{gb}, \kappa_{y'y'}^{gb}, \kappa_{z'z'}^{gb})$ is utilized for all the grain boundaries. As an example, Figure 3d shows the calculated distribution of local electric current density $J_z$ in one 2D slice of the 3D microstructure upon applying an electric field of 100 V/m along the $+z$ direction. The $J_z$ is larger at the grain boundary because the grain boundary has a higher Li-ion conductivity than the grain in this microstructure.

The effective elastic stiffness matrix $c_{klmn}^{\text{eff}}$ is calculated by $c_{klmn}^{\text{eff}} = \frac{\langle \sigma_{kl}(r)\rangle}{\varepsilon_{mn}}$ ($k,l,m,n=x,y,z$) where $\varepsilon_{mn}$ is the applied strain, and $\langle \sigma_{kl}(r)\rangle$ is the volumetric average of the local stress $\sigma_{kl}(r)$. The 3D spatial distribution of the $\sigma_{kl}(r)$ under the applied strain $\varepsilon_{mn}$ is obtained by solving the mechanical equilibrium equation,

$$\frac{\partial \sigma_{kl}}{\partial r_j} = 0, \text{ i.e., } \nabla_j[c_{klmn}(\mathbf{r})(\varepsilon_{mn} - \varepsilon_{mn}^0)] = 0 \qquad (9)$$

via the FSIP method as well [50,51]. Here $\varepsilon_{mn}^0$ is the position-dependent eigenstrain related to grains, which is set as zero herein. $c_{klmn}(\mathbf{r})$ is the local elastic stiffness coefficient in the sample coordinate system, given by,

$$c_{klmn}(\mathbf{r}) = R_{kp} R_{lq} R_{mr} R_{ns} c'_{pqrs}(\mathbf{r}) \qquad (10)$$

Where **R** is the rotation matrix, the same as that in Eq. (7); $c'_{pqrs}(\mathbf{r})$ is the local elastic stiffness matrix in the crystallographic coordinate system. Since grains have a tetragonal crystal symmetry, the $c'_{11}$, $c'_{12}$, $c'_{44}$ (expressed using the Voigt notation) are the three independent component of the elastic stiffness matrix. For voxels that are labeled as grain, $(c'_{11}, c'_{12}, c'_{44}) = (c'_{g,11}, c'_{g,12}, c'_{g,44})$. For half of the microstructures in the dataset, we set $c'_{g,11}=1.87\times10^{11}$ Pa, $c'_{g,12}=7.51\times10^{10}$ Pa, $c'_{g,44}=7.10\times10^{10}$ Pa, which are reported values for Al-doped LLZO ceramics [52]. For the other half, we set $c'_{g,11}=1.70\times10^{11}$ Pa, $c'_{g,12}=6.39\times10^{10}$ Pa, $c'_{g,44}=6.98\times10^{10}$ Pa, which are reported values for Ta-doped LLZO ceramics [52]. In voxels that are labeled as the grain boundary, $(c'_{11}, c'_{12}, c'_{44}) = (c'_{gb,11}, c'_{gb,12}, c'_{gb,44})$, where $c'_{gb,11} = 10^{r_5} c'_{g,11}$, $c'_{gb,12} = 10^{r_6} c'_{g,12}$, and $c'_{gb,44} = 10^{r_7} c'_{g,44}$. Here, $r_5$, $r_6$ and $r_7$ were randomly chosen from a uniform distribution in the range of -3 and 0 because the grain boundary, due to its disordered nature, is usually considered to have smaller elastic moduli than the grains [53].

## 4. Results

### 4.1. Training, validation, and testing of the PGNN model

The whole dataset is divided into the training dataset (4000 data points), the validation dataset (500 points) and the testing dataset (500 data points). For each epoch (that is, one complete pass of the training data points), all the weights in the PGNN model are updated using the gradient descent method through,

$$W = W - \eta \frac{\partial L}{\partial W} \qquad (11)$$

where $\eta$ is the learning rate; $L$ is the loss function, which is taken as the mean absolute error (MAE) between the predicted value $\hat{\mathbf{y}}_i$ and the true value $\mathbf{y}_i$ of the target in one batch, given by,

$$L = \frac{1}{n}\sum_{i=1}^{n}|\mathbf{y}_i - \hat{\mathbf{y}}_i|. \qquad (12)$$

Here $n$ is the batch size, which is the number of data points used in each update process through Eq. (10). Both the $\hat{\mathbf{y}}_i$ and $\mathbf{y}_i$ have three components since there are three targets: $(\kappa_{xx}^{\text{eff}}, \kappa_{yy}^{\text{eff}}, \kappa_{zz}^{\text{eff}})$ or $(c_{11}^{\text{eff}}, c_{12}^{\text{eff}}, c_{44}^{\text{eff}})$. In both the PGNN and the baseline ML models, all the input and target conductivities and elastic stiffness coefficients are expressed in the log10 scale. This is necessary because the ion conductivities ($10^{-8}\sim10^{-2}$ S/m) are small numeric values that are comparable to the error of common ML models, and because the elastic stiffness coefficients ($10^7\sim10^{12}$ Pa) are large numeric values which may cause exploding gradients due to an extremely large initial loss $L$. Since the ranges of the $\log_{10}$-scale targets ($\log_{10}\boldsymbol{\kappa}^{\text{eff}}$ and $\log_{10}\mathbf{c}^{\text{eff}}$) are still different and non-overlapping, we selected the mean absolute relative error (MARE) as the metric of property prediction performance of the PGNN and other baseline ML models, which allows us to calculate the relative difference between the predicted and the true values, *i.e.*,

$$\text{MARE} = \frac{1}{n}\sum_{i=1}^{n}\frac{|\mathbf{y}_i - \hat{\mathbf{y}}_i|}{|\mathbf{y}_i|}. \tag{13}$$

We note that the use of MARE as the performance metric and the use of MAE as the loss function are consistent with each other. Furthermore, in the PGNN model, both the node features and edge features of the input microstructure graphs are rescaled to the range of 0 to 1 using the min-max normalization, *i.e.*,

$$f' = \frac{f - f_{\min}}{f_{\max} - f_{\min}} \tag{14}$$

The $f_{\min}$ and $f_{\max}$ of each input feature $f$ are shown in Table 1. Such min-max normalization ensures that the model prediction performance will not be determined by large features. In addition, such data normalization can shorten the distance between the initial starting point and the final minimum point [54] and thereby leads to a faster model training. Indeed, our control study shows that the min-max normalization of the input features results in a faster, numerically more stable model training and significantly improves the property prediction performance.

Table 1. The minimum and maximum value of the input physical features of the PGNN model.

| Features | $f_{\min}$ | $f_{\max}$ |
|---|---|---|
| Euler angle $\alpha$ (°) | 0 | 360 |
| Euler angle $\beta$ (°) | 0 | 180 |
| Euler angle $\gamma$ (°) | 0 | 360 |
| Grain center position ($\overline{p_x}, \overline{p_y}, \overline{p_z}$) | 0 | 64 |
| Grain size (number of voxels) | 0 | $64^3$ |
| $log_{10}\boldsymbol{\kappa}$ (S/m) | -8 | -2 |
| $log_{10}\mathbf{c}$ (S/m) | 7 | 12 |

The PGNN model with different combinations of hyperparameters are independently trained using the training dataset. The model with the best performance on the validation dataset is identified as the optimized model. The hyperparameters of the optimized model are summarized in Table 2.

Table 2. Hyperparameters of the optimized PGNN model.

| | |
|---|---|
| Number of graph convolutional layer | 2 |
| Number of node-level fully connected layer | 2 |
| Number of hidden units of the fully connected layer | 1024 (1st)/128(2nd)/3(3rd) |
| Batch size | 10 |
| Learning rate | $1\times10^{-6}$ |
| Number of epochs | 50 |

The optimized model is tested independently on the testing dataset. Figure 4a shows the PGNN-predicted value vs. the true value of the $\kappa_{xx}^{\text{eff}}$, $\kappa_{yy}^{\text{eff}}$, $\kappa_{zz}^{\text{eff}}$ in log10 scale. For all three targets, the MARE values are smaller than 1.4%. These very low prediction errors demonstrate the excellent prediction ability of the PGNN model on multiple targets.

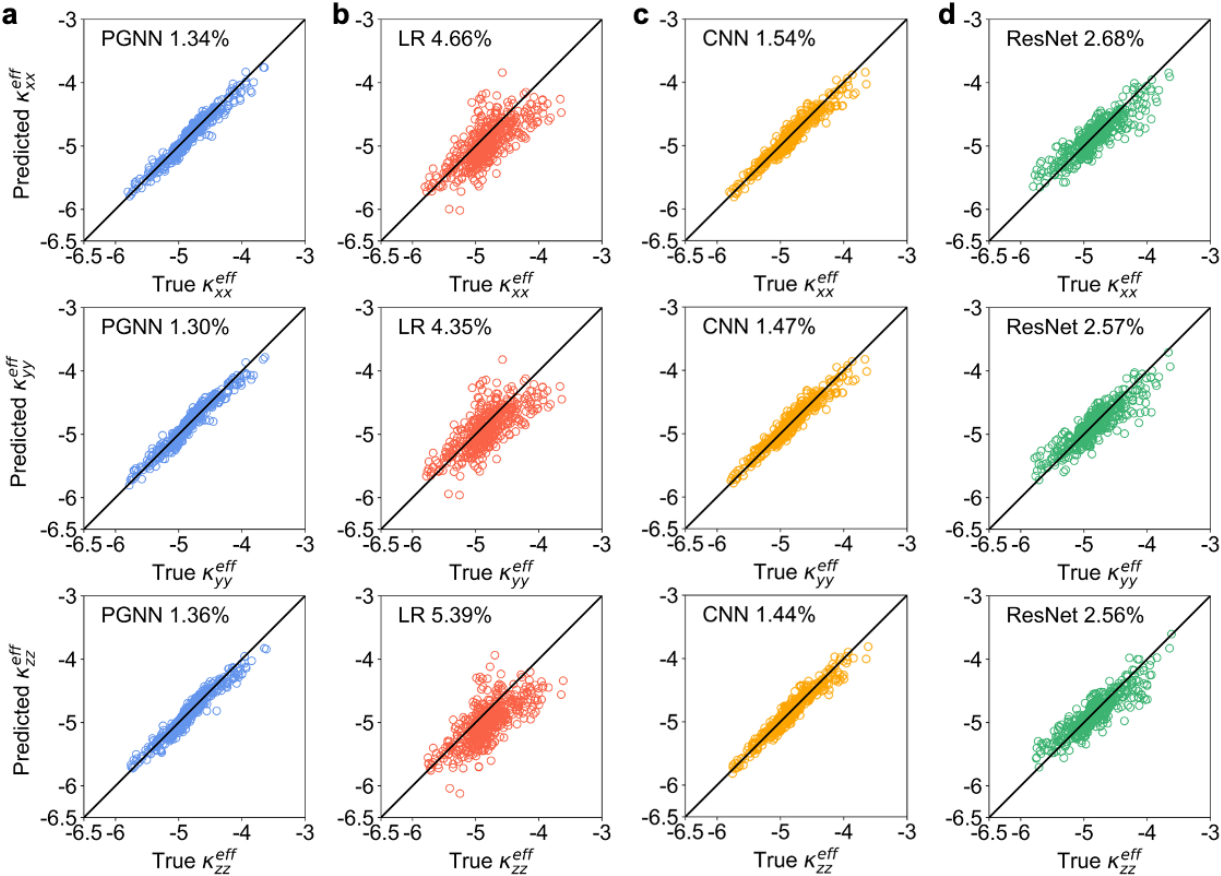

**Figure 4** The predicted values vs. the true values of the three diagonal components of the effective Li-ion conductivity matrix of the LLZO ceramic, obtained through (**a**) the PGNN model, (**b**) a linear regression (LR) model, (**c**) a CNN model, and (**d**) the ResNet model. Log10 scale is used for both the predicted and true values.

*4.2. Baseline ML models*

The MARE of the PGNN model is compared to those from three baseline ML models:

#1. A linear regression (LR) model with three fully connected layers. The numbers of hidden units of the fully connected layers are the same with those in Table 2. No activation functions are used for the fully connected layers and hence the target properties are linearly related to the input features. This simple LR model is used to determine whether there exist strong linear correlations between input features and the model output.

#2. A convolutional neural network (CNN) model, which was used to predict the multiple targets of effective elastic moduli of two-phase composites [28]. The hyperparameters of the CNN model are kept the same as in [28] except that a different number of epochs (=50) is used to get better performance on the present dataset.

#3. A ResNet model [44], which is one of the most successful general-purpose CNN models. Compared to conventional CNN models (*e.g.*, model #2) where the input of the $l+1$th convolutional layer only includes the output of the $l$th convolutional layer, ResNet model utilizes residual blocks to take the outputs of both the $l$th convolutional layer and $l+k$th convolutional layer as the input of $l+k+1$th convolutional layer, thereby establishing a so-called 'skip connection' between the $l+k+1$th convolutional layer and the $l$th convolutional layer.

The inputs of all three baseline ML models are composed of the microstructure input {**X**} and the three local ion conductivity components, which together form a matrix of 64 × 64 × 64 × 6 for a single data point. For each data point, the target properties {**Y**} are the three effective ion conductivities ($\kappa_{xx}^{\text{eff}}$, $\kappa_{yy}^{\text{eff}}$, $\kappa_{zz}^{\text{eff}}$). The CNN and ResNet model directly take the 64 × 64 × 64 × 6 matrix as the input with six channels (three Euler angles and three local conductivities). In the LR model, this 64 × 64 × 64 × 6 matrix is flattened to a 1D vector. MAE is used as the loss function in the evaluation of all three baseline ML models for consistency with the case of PGNN model.

The results show the MARE of PGNN model is the lowest in predicting all three targets. The LR model leads to the highest MARE (~5%). The property prediction performance of the PGNN model shows a statistically significant improvement of +13%, +12%, and +6% in predicting the $\kappa_{xx}^{\text{eff}}$, $\kappa_{yy}^{\text{eff}}$, $\kappa_{zz}^{\text{eff}}$, respectively, as compared to the baseline CNN model (the latter outperforms the LR and ResNet models). Furthermore, the fact that the MARE is appreciably smaller in nonlinear ML models (PGNN, CNN, and ResNet) indicates a robust nonlinear correlation between the input and outputs, which is consistent with the statistical features of the input local ion conductivity (Fig. 3c) and the target effective ion conductivity (Fig. 3c).

*4.3. Feature selection study*

We apply the sequential forward selection method [55] to quantify the relative importance of selecting individual physical features of grains and grain boundaries to improving the property prediction performance of the PGNN model. We first focus on the four types of grain (node) features (see Fig. 1d) by retaining all grain boundary (edge) features. In round one, four models were independently trained, validated, and tested using the procedures described in section 4.1, but only one of the four types of grain features is used as the node feature. As shown in Table 3, only keeping the grain conductivity leads to the smallest testing MAE. Together, the round one tests indicate that it is most important to select grain conductivity (type IV) as a node feature, second most important to select grain size (type II), next most important to select Euler angles (type III), and the least important to select grain center position (type I) as a node feature, in terms of predicting the effective ion conductivity. These findings are reasonable because the type II-IV grain features were incorporated in the physics-based simulation while the type I feature was not.

In round two, three PGNN models with grain conductivity plus one of the three other features are evaluated. It is found that the addition of grain size leads to the best model performance. In round three, two models with grain conductivity, grain size plus either the Euler angles or grain center positions are evaluated. The results show that the addition of Euler angles leads to better model performance. In round four, all four types of grain(node) features are kept, but the MAE is higher than those in round three, indicating that the grain center position is an unwanted node feature.

**Table 3.** Evaluation of the four types of grain(node) features by sequential forward selection.

| Round one | Round two | Round three | Round four |
|---|---|---|---|
| I: 0.06875 | I,IV: 0.06467 | I,II,IV: 0.0499 | I,II,III,IV: 0.06324 |
| II: 0.05596 | II,IV: 0.05713 | II,II,IV: 0.04977 | |
| III: 0.06519 | II,IV: 0.06068 | | |
| IV: 0.05271 | | | |

The four types of grain features are: grain center position (I), grain size (II), Euler angles (III), and the grain conductivities (IV), respectively, which are also shown in Fig. 1d. The numerical values are the testing MAE. Lower MAE value indicates better property prediction performance of the PGNN model.

We also evaluated the necessity of incorporating the two types of grain boundary features (four numeric values, as in Fig. 1d) as the edge features. All four types of node features are retained. It is found that keeping only the grain boundary conductivity leads to much better model performance (MAE=0.04475) compared to the case of keeping only the grain boundary thickness (MAE=0.2243). One possible reason is the grain boundary thickness, which has an identical value (=one voxel) herein, cannot uniquely represent a specific grain boundary in a specific microstructure. We expect that the grain boundary thickness would become necessary to incorporate if it shows a significant spatial variation in the dataset.

All the hyperparameters for the model training are the same as those listed in Table 2. The only difference is the model that yields the best validation MAE among the first 50 epochs (i.e., the number of epochs varies between 0 and 50) is used for testing. This early stopping technique is used to avoid overfitting, which is necessary when the lengths of the input feature vectors are different in the same round. The procedures described above based on sequential forward selection method can in future be applied to determine the necessity of incorporating other physical and morphologic features of the grain (boundary) into the node (edge) features, such as the crystallographic orientation of the grain boundaries [56], the aspect ratio of grain [39], the interface area and curvature [57] and triple line length [58] of the grain boundaries.

*4.4. Extrapolation performance of the trained PGNN Model*

To evaluate the property prediction performance of the trained PGNN model on polycrystalline microstructures that have different statistical distributions of grain size and crystallographic orientation, we generate a new dataset that contains 3D microstructures of different dimensions (64 μm$^3$ and 70 μm$^3$) by processing a large-scale 3D microstructure image obtained via electron backscatter diffraction (EBSD) [59]. As shown in Fig. 5a, we first section the original microstructure into 216 non-overlapping 3D microstructure images with an identical dimension of 64 μm × 64 μm × 64 μm. The grain label and the Euler angles of each voxel are obtained from the Dream3D file [60] associated with the published 3D EBSD data, both of which are available in Ref. [59]. The grain boundary is manually added to each microstructure based on the grain label. Specifically, If the grain label of a voxel is different from those of its neighboring voxels, the voxel will be marked as the grain boundary and re-labeled as 0. The Euler angles $\alpha$, $\beta$, $\gamma$ in the voxels that are re-labeled as grain boundary are randomly selected from the range of {0°, 360°}, {0°,180°}, {0°, 360°}, respectively. To ensure that the sectioned 3D microstructure image satisfies 3D periodic boundary condition, which is necessary for the physics-based simulations in this work, all the voxels at the surfaces of the sectioned 3D microstructure image are labeled as grain boundaries but are not shown in Fig. 5a for clarity. The same procedures were applied to generate the 216 3D microstructure images with a dimension of 70 μm × 70 μm × 70 μm. Figure 5b further shows the distributions of the grain size (top) and number of grains per microstructure image (bottom) in the cases of both 64μm³ and 70μm³, both of which are significantly different from those of the microstructures generated by Voronoi tessellation (*c.f.*, Fig. 3a). High-throughput simulations were then performed to calculate the effective ion conductivity tensors of all the 432 sectioned 3D microstructure images.

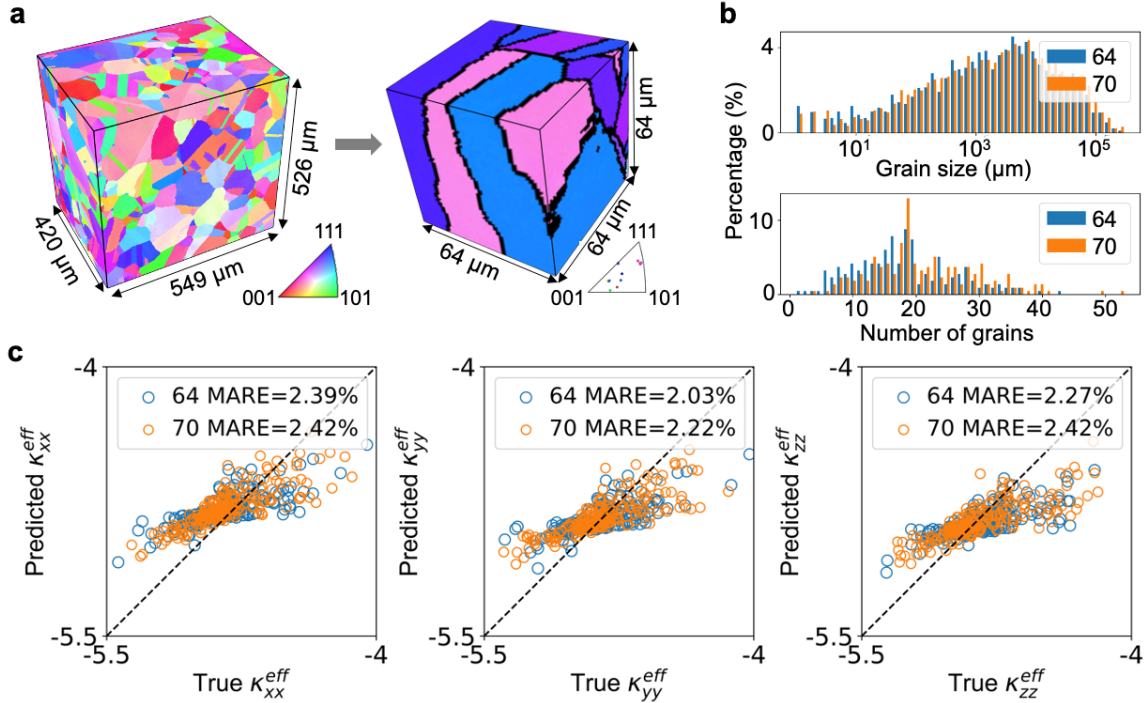

**Figure 5.** (**a**) (Left) 3D microstructure, replotted using publicly available EBSD dataset in ref. [59]; (Right) Microstructure diced from the EBSD microstructure with a size of 64 μm³ and manually added grain boundary (colored in black). (**b**) Statistical distributions of the size of all the grains and the number of grains per microstructure in all 432 diced microstructures with a size of 64 μm³ or 70 μm³. (**c**) The performance of the PGNN model trained using the Voronoi-based dataset in predicting the effective ion conductivities of the EBSD-based microstructures. Log10 scale is used for both the predicted and true values.

The PGNN model trained using the Voronoi microstructures-based dataset is directly tested on this EBSD microstructure-based dataset. Figure 5d shows the PGNN-predicted effective ion conductivities vs. the true values. Given the large difference between the Voronoi-based and the EBSD-based dataset, the MARE values increase but are still smaller than 2.4% in predicting all three targets. This result demonstrates a good extrapolation performance of our PGNN model to statistically different 3D microstructures with different dimensions.

*4.5. Transfer learning performance of the pretrained PGNN model*

The following tasks were performed to address the question: can PGNN model pretraining improve the performance of predicting a different type of effective property of the same set of microstructures?

(1) Prepare a new dataset with 604 data points of {**X**,**Y'**}, where **Y'** refer to the elastic stiffness coefficients ($c_{11}^{\text{eff}}$, $c_{12}^{\text{eff}}$, $c_{44}^{\text{eff}}$) which are the target properties for the transfer learning performance evaluation.

(2) Train, validate, and test the PGNN model using the dataset with 4000 data points of {**X**,**Y**}, where **X** are microstructures generated by Voronoi tessellation and **Y** are the effective ion conductivities. The goal of task (2) is to obtain the weights of the graph convolution layer, node-level fully connected layer, and the fully connected layer in the optimized PGNN model, and then use them as the initial weights of the transferred PGNN model.

The procedures of training, validation, and testing are the same as those described in Section 4.1, except that the target conductivities $log_{10}\kappa^{eff}$ were also normalized based on the $f_{min}$ and $f_{max}$ shown in Table 1. If directly transferring a model pretrained for predicting unnormalized $log_{10}\kappa^{eff}$ to predict unnormalized $log_{10}\mathbf{c}^{eff}$, the initial training MAE loss would be very high because the ranges of $log_{10}\kappa^{eff}$ and $log_{10}\mathbf{c}^{eff}$ are quite different.

(3) Prepare a transferred PGNN model by setting its initial weights the same as those from the pretrained PGNN model, as mentioned above. Prepare a reference PGNN model whose initial weights are randomly assigned.

(4) Divide the new dataset of 604 data points into a training dataset (80%), a validation dataset (10%), and a testing dataset (10%); independently train, validate, and evaluate the transferred and reference PGNN model using the same training dataset, during which the initial weights of both models are updated. The hyperparameters for training both the transferred and the reference model are the same as those listed in Table 2, except that a smaller number of epochs of 20 is used. This small number of epochs, together with the relatively low learning rate ($1\times10^{-6}$), ensures that the weights of the transferred model are not significantly different from the initial values.

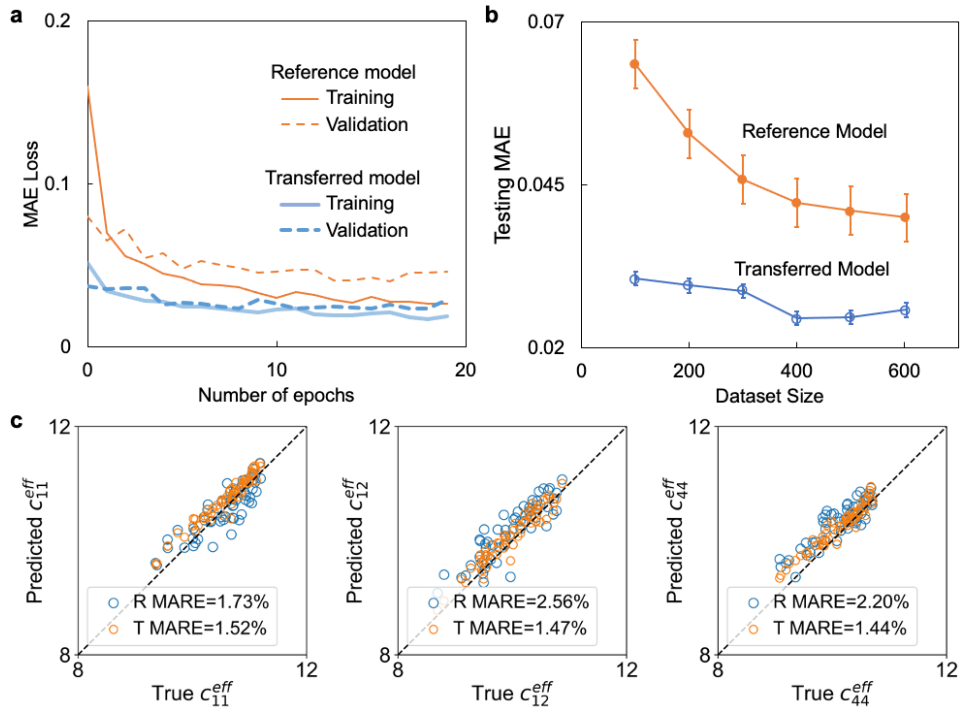

**Figure 6.** (**a**) The MAE loss curves of the transferred and reference PGNN model on the training and validation datasets. (**b**) The testing MAE of these two models as a function of the size of the training dataset. (**c**) The predicted values vs. the true values of the three targets of the testing dataset, obtained using both the transferred (T) and the reference (R) model. Log10 scale is used for both the predicted and true values.

As shown in Fig. 6a, the transferred model gives a significantly lower initial training loss and converges faster to a stable training MAE loss than the reference model. The validation loss of the transferred model remains to be much lower than that of the reference model during the entire process. The training and validation loss curves indicate that transferring the PGNN model can accelerate the model training and give more accurate prediction on new datasets. For further demonstration, a data ablation study is performed based on five sub-datasets with a size of 100, 200, 300, 400 and 500 data points that are randomly selected from the whole 604 data points. The transferred and reference PGNN models were then trained, validated, and tested on those sub-datasets. Three random number of data division are used to eliminate the influence of randomness in generating the sub-datasets. As shown in Fig. 6b, the transferred model gives lower values of testing MARE with significantly lower standard deviation than the reference model for all five sub-datasets. Figure 6c further shows the predicted values vs. the true values based on the testing dataset for both the transferred and reference model. As shown, the transferred model significantly outperforms the reference model by yielding a low testing MARE of ~1.5% for all the three targets, which is particularly remarkable considering the small size of the training dataset (604 data points). Therefore, we conclude that a PGNN model pretrained using a relatively large dataset of one property can be transferred to improve the prediction performance (higher accuracy, faster training) of another property with a smaller training dataset. This can significantly benefit the design of solid electrolytes because the throughput of measuring elastic properties is typically much lower than the experimental measurement of ion conductivities.

## 5. Conclusion

We have developed a polycrystal graph neural network (PGNN) model for predicting the effective properties of polycrystalline materials, using polycrystalline LLZO ceramic as an example. To evaluate the property prediction performance of the PGNN model, we computationally generated a large dataset of over 5000 different 3D polycrystalline microstructures with finite-width grain boundaries by Voronoi Tessellation or processing EBSD image, and then simulated the effective ion conductivities and elastic stiffness coefficients by high-throughput physics-based simulations. Using this dataset, we evaluated the performance of the PGNN in predicting all three diagonal components of the effective ion conductivity matrix. With a testing error of <1.4%, the performance of our PGNN model is superior to three baseline models including a linear regression model and two CNN models that have a relatively simple and complex structure, respectively.

We have also applied sequential forward selection to quantify the relative importance of selecting individual grain or grain boundary features on improving the property prediction accuracy, through which we have identified both the critical and unwanted node/edge features in our PGNN model. Additionally, we show that the trained PGNN model shows an acceptable testing error of <2.4% on a new dataset where the 3D polycrystalline microstructures have completely different statistical features, demonstrating a good extrapolation performance. Furthermore, compared to a PGNN model trained from scratch, the PGNN model pretrained using the dataset of microstructure-conductivity yields a faster training speed and a lower error in predicting the effective elastic stiffness coefficients of the same set of microstructures. The property prediction performance of the current PGNN model could be further improved by optimizing the update function to better approximate the physical interactions among the local structure elements for building a physics-informed neural network [61–63]. Finally, we would like to remark that the graph-based presentation can in principle be extended to any other mesoscopic systems that contain a set of mutually interacting elements, such as a group of colloidal nanoparticles in liquid solutions, mechanically interacting particles in a composite battery cathode [64], electronic [65] or magnetic [66] quantum-dot cellular automata, and biological cells, thereby creating mesoscale graph inputs for a wider variety of downstream tasks (e.g., particle tracking [67], and prediction of the moving trajectories of particles [68,69]) in addition to the effective property prediction.

**Acknowledgements**

Acknowledgment is made to the donors of The American Chemical Society Petroleum Research Fund for partial support of this research, under the award PRF # 61594-DNI9 (M.D. and J.-M.H.). Xuanhan Liu's participation was assisted by the National Science Foundation (NSF) under grant OAC 2017072. Any opinions, findings, and conclusions or recommendations expressed in this publication are those of the author(s) and do not necessarily reflect the views of the NSF. The simulations were performed using Bridges at the Pittsburgh Supercomputing Center through allocation TG-DMR180076 from the Advanced Cyberinfrastructure Coordination Ecosystem: Services & Support (ACCESS) program, which is supported by NSF grants #2138259, #2138286, #2138307, #2137603, and #2138296.

**Dataset availability**

Datasets of the microstructure-effective Li-on conductivity and the microstructure-effective elastic stiffness coefficients of polycrystalline LLZO ceramics are available in GitHub via https://github.com/mdai26/PGNN.

**Code availability**

The in-house codes of polycrystalline microstructure generation through Voronoi Tessellation and the PGNN model can both be accessed in GitHub via https://github.com/mdai26/PGNN.